# AN AUTONOMOUS PASSIVE NAVIGATION METHOD FOR NANOSATELLITE EXPLORATION OF THE ASTEROID BELT

Leonard Vance,[*] Jekan Thangavelautham,[†] and Erik Asphaug[‡]

There are more than 750,000 asteroids identified in the main belt. These asteroids are diverse in composition and size. Some of these asteroids can be traced back to the early solar system and can provide insight into the origins of the solar system, origins of Earth and origins of life. Apart from being important targets for science exploration, asteroids are strategically placed due to their low-gravity well, making it low-cost to transport material onto and way from them. They hold valuable resources such as water, carbon, metals including iron, nickel and platinum to name a few. These resources maybe used in refueling depots for interplanetary spacecraft and construction material for future space colonies, communication relays and space telescopes. The costs of getting to the main asteroid belt, combined with large numbers of objects to be explored encourage the application of small spacecraft swarms. The size and capability of the resulting nano-spacecraft can make detection from Earth difficult. This paper discusses a method by which a spacecraft can establish ephemeris autonomously using line of sight measurements to nearby asteroids with Extended Kalman Filtering techniques, without knowing accurate ephemeris of either the asteroids or the spacecraft initially. A description of the filter implementation is followed by examples of resultant performance.

**INTRODUCTION**

The number of asteroids currently cataloged exceeds seven-hundred fifty thousand. These asteroids are known to be diverse in size, composition and origin. Asteroids surfaces undergo weathering and are covered by dust that hide their internal composition. Older asteroids are thought to undergo the 'Brazil Nut' effect, where large boulders settle to the top, leaving a different composition of material below. Ground observation and even in-space observation cannot be used to guarantee what is beneath these top layers. Systematically performing close-up surveys, sample analysis/return and use of bistatic radar is perhaps the only sure-way of determining the internal composition of these asteroids. The daunting task of exploring any statistically significant subset of this total will likely involve swarms of small spacecraft, which would place a challenging requirement on any terrestrial based navigation scheme implemented. A technique for some type of autonomous navigation would be useful, allowing spacecraft to navigate within the asteroid belt without support from an Earth based system.

---

[*] PhD Candidate, Aerospace and Mechanical Engineering, Univ. of Arizona, 1130 N Mountain, Tucson, AZ 85721
[†] Assistant. Professor, Aerospace and Mechanical Engineering, Univ. of Arizona, 1130 N Mountain, Tucson, AZ 85721
[‡] Professor. Lunar and Planetary Laboratory, Univ. of Arizona, 1629 E University, Tucson AZ 85721

Navigation techniques using passive line-of-sight measurements are well known within the community. The ability to infer ephemeris information of an orbital object by tracking it from a known position is well understood, as is the ability to infer own ephemeris by tracking an object with known position and velocity. This paper shows that it is also possible, however, to infer both own-ship and tracked-object ephemeris given knowledge of the gravitational influences both are subjected to.

Figure 1 shows the general nature of the problem. It is well established that position and velocity data for an overhead satellite can be extracted from passive line of sight measurements from the ground. Observability of range is inferred from the gravitational attraction of the orbiting object. Likewise, as a simple extension, it is also possible to do the same job from an orbiting satellite. If you know your own position and velocity, you can infer the position and velocity of another satellite given a time history line of sight measurements.

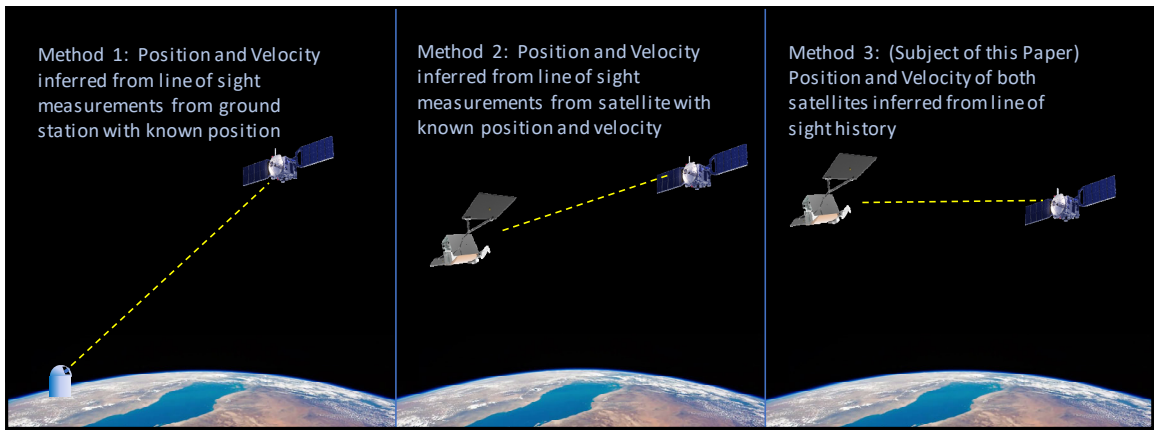

Figure 1: Auto-navigation using other objects under gravitational influence is the logical descendent of established navigation techniques

With some additional thought, it is possible that the position and velocity of both the target and ownship satellites can be inferred from a single line of sight history. With some exceptions, it can be asserted that any line of sight history between two objects gravitationally influenced by a common source is the result of only one specific target and home satellite trajectory. Furthermore, the tracking of more than one object improves navigational performance, and this paper evaluates the utility of tracking a $2^{nd}$ object (an asteroid in this case).

**RELATED WORKS**

Current missions beyond the Earth-Moon system utilize the Deep Space Network or equivalent network of ground stations to determine position in space and perform navigation through Doppler ranging. As more small-space missions are planned and many active spacecraft in operation, the DSN will be under strain and there will be an important need to find viable alternatives. A substantial body of work exists regarding deep space auto-navigation. The most popular methodology is triangulation or inferential estimation by recording the line of sight history of a known object from the spacecraft. Others propose the precise measurement of pulsar time histories, and some propose range measurements to crosslink spacecraft or optical measurement of known celestial objects combined with doppler measurements of solar output. None of these techniques propose utilizing passive line of sight angular tracking to objects with poor position knowledge.

Multiple references are available which explore the ability to infer ownship position and velocity via line of sight history of an object with well known location. S. Bhaskaran et al.[4], Karimi et al.[7], Polle et al.[8], and Riedel et al[9] all propose some form or combination of this technique, tracking one or more objects with known position. K. Hill and G. H. Born[5] propose range measurements to cross link vehicles with known position, while Yim et al.[3] combines angular measurements to the earth and spectrometer measurements of doppler shift from the sun. Finally, a body of work represented by Shemar et al.[10] explore the possibility of using x-ray pulsars, measuring the arrival of pulses and thus inferring distance from them. This approach would then work much like GPS, triangulating using the change in linear distances to 3 or more neutron stars to establish position.

**METHOD**

The Extended Kalman Filter (EKF) is a well-known, pseudo-optimal predictor-corrector filter widely used for a wide variety of aerospace purposes. This paper shows the adaptation of the basic filter structure for this particular problem. The established form of the EKF[*] starts by propagating the current estimate and covariance matrices forward by one timestep.

$$\hat{x}(k+1|k) = \hat{x}(k|k) + \int_{t_k}^{t_{k+1}} f[\hat{x}(t|t_k), u^*(t), t] dt \qquad (1)$$

$$P(k+1|k) = \Phi(k+1,k) P(k|k) \Phi'(k+1,k) + Q_d(k) \qquad (2)$$

This followed by calculation of the Kalman gains:

$$K(k+1) = P(k+1|k)[H_x'(k+1) P(k+1|k) H_x'(k+1) + R(k+1)]^{-1} \qquad (3)$$

The state vector estimate can then be updated:

$$\begin{aligned}\hat{x}(k+1|k+1) = \hat{x}(k+1|k) + K(k+1)\{z(k+1) \\ -h[\hat{x}(k+1|k), u^*(k+1), k+1] - H_u(k+1)\partial u(k+1)\}\end{aligned} \qquad (4)$$

Finally, also update the covariance matrix:

$$P(k+1|k+1) = [I - K(k+1) H_x(k+1)] P(k+1|k) \qquad (5)$$

The definition of these parameters are consistent with normal EKF usage:

$\hat{x}$ = state estimate
$k$ = current time step
$k+1$ = next time step
$f$ = function propagating state estimate $\hat{x}$ in time
$t$ = time
$u^*$ = nominal input
$P$ = Covariance Matrix
$\Phi$ = Linearized time propagation matrix
$Q_d$ = State error propagation matrix
$K$ = Kalman gain matrix
$h$ = function taking state vector to measurements
$H_x$ = Jacobean of $h$ w/r to state vector x
$H_u$ = Jacobean of $h$ w/r to external inputs
$I$ = Identity matrix

---

[*] "Lessons in Digital Estimation theory", Jerry M. Mendel, Prentice-Hall, 1987

The function u* and its derivatives are zero for the purposes of this exercise since there are no external forces acting on the system, and as such, (1) through (5) can be simplified to give

$$\hat{x}(k+1|k) = \hat{x}(k|k) + \int_{t_k}^{t_{k+1}} f[\hat{x}(t|t_k), t]dt \qquad (6)$$

$$P(k+1|k) = \Phi(k+1,k)P(k|k)\Phi'(k+1,k) + Q_d(k) \qquad (7)$$

for the propagation equations, and

$$K(k+1) = P(k+1|k)[H'_x(k+1)P(k+1|k)H'_x(k+1) + R(k+1)]^{-1} \qquad (8)$$

$$\hat{x}(k+1|k+1) = \hat{x}(k+1|k) + K(k+1)\{z(k+1) - h[\hat{x}(k+1|k)]\} \qquad (9)$$

$$P(k+1|k+1) = [I - K(k+1)H_x(k+1)]P(k+1|k) \qquad (10)$$

for the update equations.

The state estimation vector $\hat{x}$ contains 3 element position, velocity and accelerations for each object. Since this paper discusses tracking two objects from a home vehicle, there are 27 elements in the state estimation vector comprising:

$$[\hat{x}]_{(27\times1)} = \begin{bmatrix} \hat{x}_T \\ \hat{x}_1 \\ \hat{x}_2 \end{bmatrix} \begin{array}{l} \text{(9x1)}\\ \text{each} \end{array} \quad \hat{x}_T \begin{cases} \hat{r}_T \text{ (3x1)} \begin{cases} x \text{ position estimate, tracking vehicle} \\ y \text{ position estimate, tracking vehicle} \\ z \text{ position estimate, tracking vehicle} \end{cases} \\ \hat{v}_T - x,y,z \text{ velocity estimates, tracking vehicle} \\ \hat{a}_T - x,y,z \text{ acceleration estimates, tracking vehicle} \end{cases}$$
$\hat{x}_1$ Object 1 estimates
$\hat{x}_2$ Object 2 estimates

The function $f$ is a simple orbital propagator, utilizing the basic gravitational law

$$a = -\frac{Gm_{sun}r}{|r|^3} \qquad (11)$$

where the $r$'s are taken from the position estimates in the state vector. The corresponding matrix $\Phi$ is the Jacobean of this with respect to $\hat{x}$, giving a 27 × 27 matrix. The function $h$ takes state variables and constructs the measurements $z$, and is therefore of the form:

$$z = h(\hat{x}) \qquad (12)$$

Since the measurements for this system are the normalized line of sight vectors from the tracking vehicle to the two asteroids, the resulting function $h$ is

$$h = \begin{bmatrix} \dfrac{x_1 - x_T}{\sqrt{(x_1 - x_t)^2 + (y_1 - y_t)^2 + (z_1 - z_t)^2}} \\ \dfrac{y_1 - y_T}{\sqrt{(x_1 - x_t)^2 + (y_1 - y_t)^2 + (z_1 - z_t)^2}} \\ \dfrac{z - z_T}{\sqrt{(x_1 - x_t)^2 + (y_1 - y_t)^2 + (z_1 - z_t)^2}} \\ \dfrac{x_2 - x_T}{\sqrt{(x_2 - x_t)^2 + (y_2 - y_t)^2 + (z_2 - z_t)^2}} \\ \dfrac{y_2 - y_T}{\sqrt{(x_2 - x_t)^2 + (y_2 - y_t)^2 + (z_2 - z_t)^2}} \\ \dfrac{y_2 - y_T}{\sqrt{(x_2 - x_t)^2 + (y_2 - y_t)^2 + (z_2 - z_t)^2}} \end{bmatrix} \quad (13)$$

The Jacobian of this function with respect to each component of the state estimate vector $\hat{x}$ provides the resulting $6 \times 27$ $H_x$ matrix.

The $R$ matrix represents the system measurement noise, and this takes the form of angular uncertainties along the line of sight to objects 1 and 2. The measurement vector $z$ is the two normalized line of sight vectors to each target as given by (13), so given an angular uncertainty $\varepsilon$ in each axis, the covariance matrix for measurement errors to object 1 in line of sight coordinates is:

$$R_{1,los} = \begin{bmatrix} 1 & 0 & 0 \\ 0 & \varepsilon^2 & 0 \\ 0 & 0 & \varepsilon^2 \end{bmatrix} \quad (12)$$

This must then be rotated into the heliocentric coordinate frame used by the rest of the filter. Derivation of the coordinate transform matrix is done by establishing sequential coordinate frame rotation angles about the $z$ and then $y$-axes respectively, providing a coordinate transformation matrix of

$$T_{I,los1} = \begin{bmatrix} \cos\psi_1 \cos\theta_1 & \sin\psi_1 \cos\theta_1 & -\sin\theta_1 \\ -\sin\psi_1 & \cos\psi_1 & 0 \\ \cos\psi_1 \sin\theta_1 & \sin\psi_1 \sin\theta_1 & \cos\theta_1 \end{bmatrix} \quad (13)$$

Where the sequential angles of rotation about z and y axes respectively are:

$$\psi_1 = \tan^{-1}\left(\frac{y_1 - y_T}{x_1 - x_T}\right) \quad (14)$$

$$\theta_1 = \tan^{-1}\left(\frac{z_1 - z_T}{\sqrt{(x_1 - x_T)^2 + (y_1 - y_T)^2}}\right) \quad (15)$$

This matrix takes a vector from inertial to the line of sight coordinated frame with respect to object 1. Using a similarity transformation to convert from line of sight to inertial coordinates, we have

$$R_{1,inertial} = T_{I,los1} R_{1,los} T_{I,los1}^T \quad (16)$$

The second **R** matrix to the 2$^{nd}$ object can be similarly calculated, and the two can be combined to provide the overall $6 \times 6$ input to the filter equations.

$$R = \begin{bmatrix} R_{1,inertial} & 0 \\ 0 & R_{2,inertial} \end{bmatrix} \quad (17)$$

Finally, the 27 × 27 state process noise matrix **Q** is constructed with arbitrarily small diagonal elements for the purpose of establishing feasibility.

$$Q = 1.0 x 10^{-30} * I_{27x27} \quad (18)$$

**RESULTS**

The implementation of the above filter is tested in an environment derived from the JPL asteroid ephemeris database[*]. The tracking vehicle is established in a circular orbit at the median heliocentric altitude, and a collection of nearby asteroids is established to provide a pool of possible objects to track and test filter performance. The test case chosen utilizes asteroids with catalog numbers 32,675 and 272,344, and an argument of latitude of $\pi/3$ for the tracking spacecraft. The resulting geometry is shown in Figure 2, along with a history of the distances between the tracking spacecraft and the two asteroids. The simulation is then executed for 1 year, sampling once per hour. The standard deviation of the line of sight error is 10 microradians in vertical and horizontal axes. Initial position and velocity variances are set to $(100,000km)^2$ and $(1 km/s)^2$ respectively.

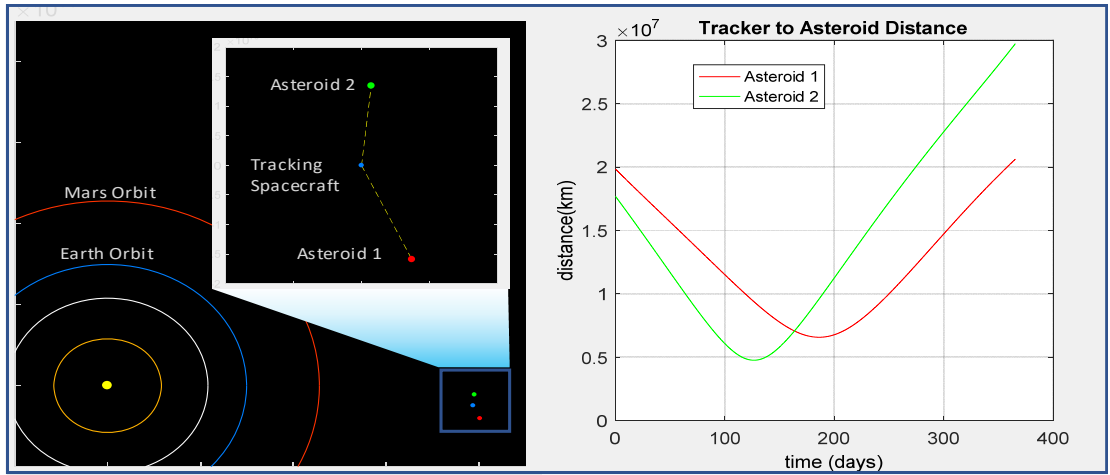

Figure 2: Test Geometry Overview

**Case 1: Single object tracking**

This case uses only tracking information from asteroid 1 to establish position and velocity. Results are shown in Figure 3, illustrating good convergence in position and velocity in all 3 axes. The associated position and velocity errors for the asteroid tracking are then summarized similarly in Figure 4. Results are similar to those seen for the observer spacecraft, plausibly because the problem is nearly symmetric given the identical starting covariances.

An overall assessment of filter performance assuming the results actually converge can be extracted from the time history of covariance magnitudes as shown in Figure 5. The convergence of the observer and asteroid are nearly identical given the initial condition stated, and steady im-provement in position and velocity knowledge continue well after the closest point of approach of the tracked asteroid.

---

[*] NASA/JPL website for asteroid orbital ephemeris. https://ssd.jpl.nasa.gov/?sb_elem

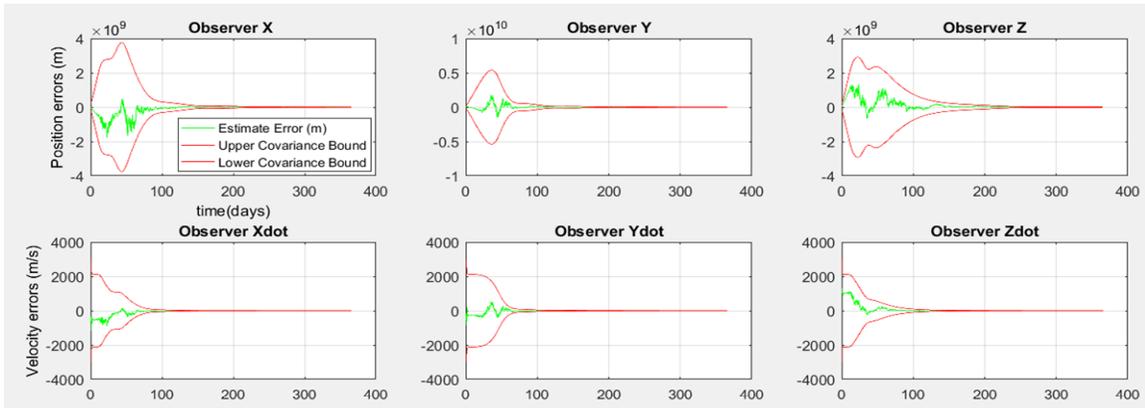

Figure 3: Case 1 (single asteroid track) residual error and 3-sigma covariance bounds for observer position and velocity errors.

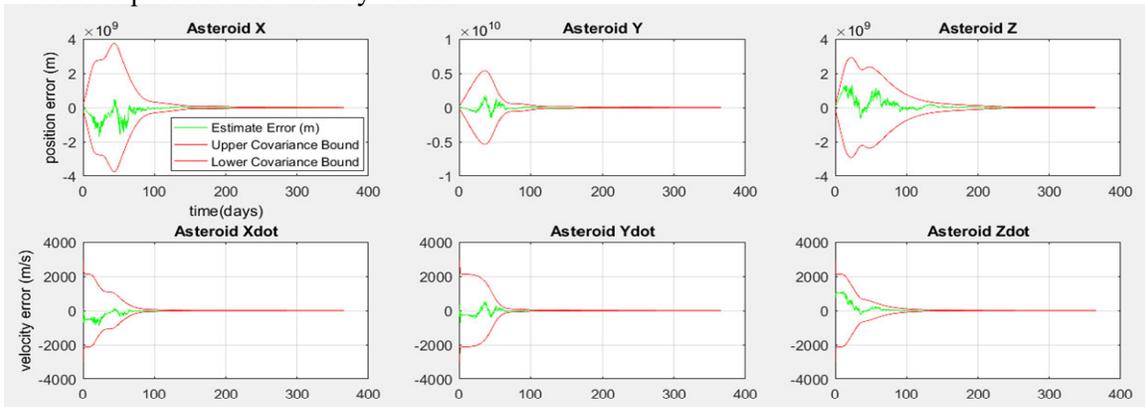

Figure 4: Case 1 (single asteroid track) residual error and 3-sigma covariance bounds for asteroid position and velocity errors.

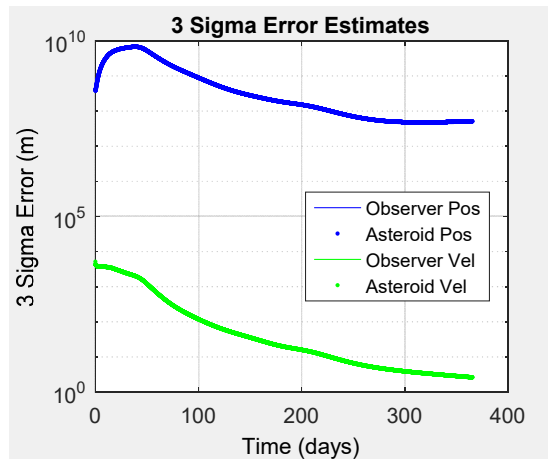

Figure 5: Case 1 error bound convergence

## Case 2: Two Asteroids Tracked

Case 2 uses the same first asteroid as Case 1 but adds a 2nd asteroid as shown in Figure 2. Once again, all objects are given the same initial covariances outlined in the previous section and the simulation is then executed over a 1-year epoch with both asteroids being used to update the EKF. Basic results for the observer spacecraft are shown in Figure 6, and these show a significant improvement in performance over the single asteroid case shown in Figure 3, with noticeable reductions in the peak errors, and earlier convergence seen in each axis. Overall results for both the satellite and the two asteroids are given in Figure 7, noting that overall performance improves by a factor of 2 to 5.

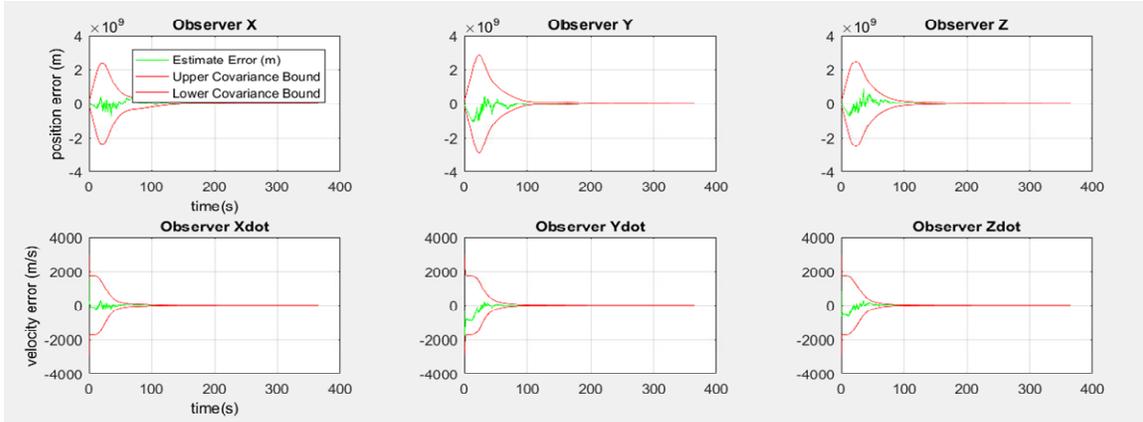

figure 6: Case 2 (2 asteroids tracked) residual error and 3-sigma covariance bounds for observer are significantly improved from single asteroid

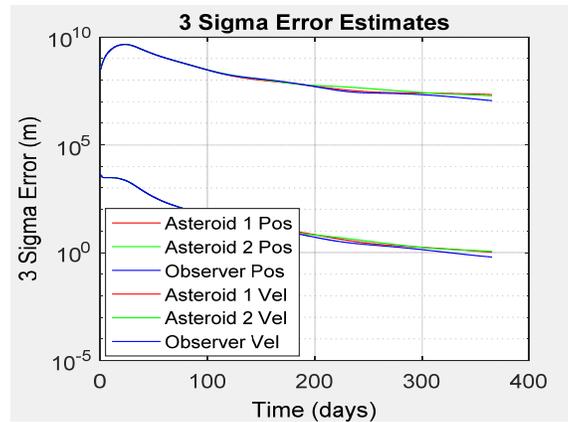

Figure 7: Case 2 error bounds converge faster because two asteroids are being tracked

## CONCLUSIONS

The concept of achieving auto-navigation using passive line of sight measurements from poorly known objects under the influence of a common gravitational field is demonstrated and quantified, with results showing convergence for both the observer and asteroid position and velocity estimates. Significant improvements in performance are observed when the number of tracked asteroids is increased from one to two, and it is expected that the addition of more observations using other asteroids would improve observability and therefore filter performance even further. The relatively low tracking rate required for this technique allows the possibility of perhaps dozens of asteroids being tracked, with the expectation of significantly improved accuracy.

Mission design for future swarm asteroid exploration missions can incorporate this auto-navigation technique to offload Earth-based navigation assets such as the DSN by providing an inexpensive alternative. In addition, the ephemeris of the tracked asteroids is improved almost as much as that of the observer, and this information can be used to update the orbital parameters of every asteroid tracked.